\documentclass[%
 reprint,
superscriptaddress,
nofootinbib,
longbibliography,
bibnotes,
 amsmath,amssymb,
 aps,
 pre,
]{revtex4-1}

\usepackage{graphicx}
\usepackage{dcolumn}
\usepackage{bm}
\usepackage{color}
\definecolor{OliveGreen}{rgb}{0,0.6,0}
\definecolor{auburn}{rgb}{0.43, 0.21, 0.1}
\definecolor{blue_violet}{rgb}{0.54, 0.17, 0.89}
\definecolor{hokie_maroon}{RGB}{99, 0, 49} 
\definecolor{hokie_orange}{RGB}{207, 69, 32} 

\usepackage{appendix}
\usepackage{chngcntr}
\usepackage{etoolbox}
\usepackage{lipsum}
\usepackage{mathtools}

\AtBeginEnvironment{appendices}{%
\addcontentsline{toc}{section}{Appendices}
\counterwithin{figure}{section}
\counterwithin{equation}{section}
\counterwithin{table}{section}
}

\begin{document}


\title{First-principles calculation of electronic density of states and Seebeck coefficient in transition-metal-doped Si--Ge alloys}

\author{Ryo Yamada}
\email{r-yamada@mat.eng.osaka-u.ac.jp}
\affiliation{Division of Materials and Manufacturing Science, Graduate School of Engineering, Osaka University, Suita, Osaka 565-0871, Japan}
\author{Akira Masago}
\affiliation{Center for Spintronics Research Network (CSRN), Osaka University, Toyonaka, Osaka 560-8531, Japan}
\author{Tetsuya Fukushima}
\affiliation{Institute for Solid State Physics, University of Tokyo, Kashiwa, Chiba 277-8581, Japan}
\author{Hikari Shinya}
\affiliation{Center for Spintronics Research Network (CSRN), Osaka University, Toyonaka, Osaka 560-8531, Japan}
\affiliation{Research Institute of Electrical Communication, Tohoku University, Sendai, Miyagi 980-8577, Japan}
\author{Tien Quang Nguyen}
\affiliation{Institute for NanoScience Design, Osaka University, Toyonaka, Osaka 560-8531, Japan}
\author{Kazunori Sato}
\affiliation{Division of Materials and Manufacturing Science, Graduate School of Engineering, Osaka University, Suita, Osaka 565-0871, Japan}
\affiliation{Center for Spintronics Research Network (CSRN), Osaka University, Toyonaka, Osaka 560-8531, Japan}

\date{\today}

\begin{abstract}
High $ZT$ value and large Seebeck coefficient have been reported in the nanostructured Fe-doped Si--Ge alloys. In this work, the large Seebeck coefficient in Fe-doped Si--Ge systems is qualitatively reproduced from the computed electronic density of states, where a hybrid functional, HSE06, is used for an exchange-correlation functional, as well as a special quasi-random structure (SQS) for a disordered atomic configuration. Furthermore, by replacing Fe with other transition metals, such as Mn, Co, Ni, Cu, Zn, and Au, a better dopant that produces a larger Seebeck coefficient in Si--Ge alloy systems is explored. 
\end{abstract}

\pacs{Valid PACS appear here}
\maketitle

A vast amount of available energy has been wasted as heats, and it is expected that thermoelectric materials are employed to extract electricity from wasted-heats. Si--Ge alloys are known as one of the cheapest nontoxic thermoelectric materials utilized at high temperatures. The dimensionless figure of merit, $ZT$, of Si--Ge alloys, however, is quite small;  $ZT<1$ for both $p$- and $n$-type thermoelectric materials \cite{rowe2018crc}. 

The small $ZT$ values in Si--Ge alloys have been somewhat improved with the use of a nanostructuring approach, where a phonon conductivity is reduced by making a grain size small. To further increase the $ZT$ values of Si--Ge alloys, there are some attempts to modify their electronic band structure by doping transition metals \cite{delime2019large,omprakash2019au,takeuchi2019private}, and it has been reported that a quite high $ZT$ value, $ZT>1.88$ (at $T=873$\;K), was obtained in the nanostructured Si$_{0.55}$Ge$_{0.35}$P$_{0.10}$Fe$_{0.01}$ sample, as well as a large Seebeck coefficient, $|S|>517$\;$\mu$V/K (at $T=673$\;K) \cite{delime2019large}. It is believed that the large Seebeck coefficient originated from a strong peak at the edge of the conduction band in the electronic density of states generated by the Fe-doping (a so-called 3$d$ impurity state), and this large Seebeck coefficient increased the $ZT$ value through the relation, $ZT \propto S^2$. Although high $ZT$ values as well as large Seebeck coefficients have also been observed in other transition-metal-doped Si--Ge alloys, such as Au- \cite{omprakash2019au} and Ni-doped systems \cite{takeuchi2019private}, their $ZT$ values and Seebeck coefficients were not as high as those of the Fe-doped system.

Although an occurrence of the 3$d$ impurity states have been confirmed in an Fe-doped Si system using an electronic band structure calculation \cite{delime2019large}, that in Fe-doped Si--Ge system has not been confirmed yet either from an experimental or theoretical approach. In this work, therefore, the electronic density of states in Fe-doped Si--Ge alloys is calculated using an electronic band structure calculation, and the reported large Seebeck coefficient is reproduced from the computed electronic density of states. In addition, by substituting Fe with other transition metal (TM=Mn, Co, Ni, Cu, Zn, or Au), a better dopant for Si--Ge alloys that produces a larger Seebeck coefficient than that of Fe-doped systems is sought.



Si--Ge alloy forms a single solid solution over the whole composition range \cite{olesinski1984ge}, and it has been reported that the nanostructured Si$_{0.55}$Ge$_{0.35}$P$_{0.10}$Fe$_{0.01}$ sample sintered at 873\;K and 400\;GPa was also composed of a single solid solution \cite{delime2019large}. To describe a disordered configuration in a solid solution, a special quasi-random structure (SQS) \cite{zunger1990special,wei1990electronic} is employed here, which is the best periodic supercell to mimic the true disordered configuration using a small number of particles. 

A SQS, which contains 64 atoms with a diamond structure ($2 \times 2 \times2 $), is searched using {\it mcsqs} code available in the Alloy Theoretic Automated Toolkit (ATAT) \cite{van2013efficient}. The following criterion is used to determine the SQS; pair correlation functions of SQS become nearly identical to those of the random alloy up to the third-nearest neighbors. The following alloy compositions are considered here; Si$_{0.500}$Ge$_{0.484}$Fe$_{0.016}$ (or Si$_{32}$Ge$_{31}$Fe$_{1}$) and Si$_{0.781}$Ge$_{0.203}$Fe$_{0.016}$ (or Si$_{50}$Ge$_{13}$Fe$_{1}$) for Fe-doped systems, and Si$_{0.500}$Ge$_{0.484}$TM$_{0.016}$ (or Si$_{32}$Ge$_{31}$TM$_{1}$) for TM-doped ones, where TM=Mn, Co, Ni, Cu, Zn, and Au from either 3$d$ or 5$d$ transition metals. 

The electronic density of states of the SQSs is calculated using the projector augmented-wave (PAW) method \cite{kresse1996efficiency} as implemented in the Vienna $Ab$ $Initio$ Simulation Package (VASP). Since it was confirmed that magnetic moments disappear during the electronic self-consistent-loop for all the transition-metal-doped systems, a non-spin-polarization calculation is conducted in this work. As an exchange-correlation functional, a hybrid functional introduced by Heyd, Scuseria, and Ernzerhof (HSE) \cite{heyd2003hybrid} is employed, setting the range-separation parameter to 0.207\;\AA$^{-1}$ (known as HSE06 \cite{krukau2006influence}). The total energy of the supercell is minimized in terms of the volume, and their atomic positions are optimized until all force components are smaller than 0.01\;$\mathrm{eV/\AA}$. The plane wave cut-off energy is set to 350\;eV, and the integration over the Brillouin zone is done, using Gaussian smearing of 0.05\;eV with $2 \times 2 \times 2$ and $6 \times 6 \times 6$ $k$-points for the structure relaxation and density of states calculations, respectively. Note that the equilibrium volume of the non-doped system is used for those of the doped systems, assuming that the volume change by doping is negligible; i.e., the equilibrium volume of Si$_{0.500}$Ge$_{0.500}$ (or Si$_{32}$Ge$_{32}$) and Si$_{0.781}$Ge$_{0.219}$ (or Si$_{50}$Ge$_{14}$) are used for those of Si$_{0.500}$Ge$_{0.484}$Fe$_{0.016}$ (Si$_{0.500}$Ge$_{0.484}$TM$_{0.016}$) and Si$_{0.781}$Ge$_{0.203}$Fe$_{0.016}$ systems, respectively.

It is noteworthy that because a Seebeck coefficient is quite sensitive to the magnitude of a band gap, the use of a hybrid functional (HSE06) is important to evaluate the Seebeck coefficient in the doped Si--Ge alloys reliably. The band gap in the Si--Ge systems, as well as lattice constant, calculated using the HSE06 and generalized gradient approximation (GGA) of Perdew-Burke-Ernzerhof (PBE) \cite{perdew1996generalized} are compared in supplementary data. It is shown that HSE06 can reliably estimate both the band gap and lattice constant of the Si--Ge system compared to the GGA/PBE functional (see figures in supplementary data).

The Seebeck coefficient, $S$, is defined as a constant of proportionality between the electric field, $\bold{E}$, and temperature gradient, $\Delta T$; i.e., $\bold{E} = S \Delta T$. The Seebeck coefficient is given from the linear response theory as \cite{mott1936theory}
\begin{equation}
\begin{split}
S (T) = -\frac{1}{|e| T} \frac{ \int_{-\infty}^{\infty} \sigma (\epsilon , T) (\epsilon - \mu) \left( \frac{\partial f_{\mbox{\scriptsize FD}}(\epsilon , T)}{\partial \epsilon}\right) d\epsilon }{ \int_{-\infty}^{\infty} \sigma (\epsilon , T) \left( \frac{\partial f_{\mbox{\scriptsize FD}}(\epsilon , T)}{\partial \epsilon}\right) d\epsilon }  \; , \label{eq:seebeck_original}
\end{split}
\end{equation}
where $e$ is the unit charge of electron, $\mu$ is the chemical potential, $\epsilon$ is the energy, $f_{\mbox{\scriptsize FD}}(\epsilon , T)$ is the Fermi-Dirac distribution, and $\sigma (\epsilon , T)$ is the spectral conductivity. From the Bloch--Boltzmann theory, the spectral conductivity for an isotropic material is written as 
\begin{equation}
\begin{split}
\sigma (\epsilon , T) = \frac{e^2}{3} D(\epsilon) v^2(\epsilon) \tau(\epsilon, T) \; , \label{eq:spectral_conductivity}
\end{split}
\end{equation}
where $D(\epsilon)$ is the electronic density of states, $v (\epsilon)$ is the group velocity, and $\tau(\epsilon, T)$ is the relaxation time. 

For a nanostructured bulk sample, an electron mean-free path, $l$ ($=v \tau$), can be approximated to a nanograin size, $a$ (i.e., $v \tau \approx a$) (a so-called small-grain-size limit \cite{bera2010thermoelectric,hao2016high}). Then, the Seebeck coefficient, Eq.\;\eqref{eq:seebeck_original}, becomes
\begin{equation}
\begin{split}
S (T) \approx -\frac{1}{|e| T} \frac{ \int_{-\infty}^{\infty} D (\epsilon) v (\epsilon) (\epsilon - \mu) \left( \frac{\partial f_{\mbox{\scriptsize FD}}(\epsilon , T)}{\partial \epsilon}\right) d\epsilon }{ \int_{-\infty}^{\infty} D (\epsilon) v (\epsilon) \left( \frac{\partial f_{\mbox{\scriptsize FD}}(\epsilon , T)}{\partial \epsilon}\right) d\epsilon } \; . \label{eq:seebeck_approximate0}
\end{split}
\end{equation}
Furthermore, by assuming that the group velocity, $v(\epsilon)$, is not sensitive to energy (i.e., $v (\epsilon) \approx v$), Eq.\;\eqref{eq:seebeck_approximate0} can be simplified to 
\begin{equation}
\begin{split}
S (T) \approx -\frac{1}{|e| T} \frac{ \int_{-\infty}^{\infty} D (\epsilon) (\epsilon - \mu) \left( \frac{\partial f_{\mbox{\scriptsize FD}}(\epsilon , T)}{\partial \epsilon}\right) d\epsilon }{ \int_{-\infty}^{\infty} D (\epsilon) \left( \frac{\partial f_{\mbox{\scriptsize FD}}(\epsilon , T)}{\partial \epsilon}\right) d\epsilon } \; . \label{eq:seebeck_approximate}
\end{split}
\end{equation}
From Eq.\;\eqref{eq:seebeck_approximate}, the Seebeck coefficient for a nanostructured bulk sample can be estimated just from an electronic density of states. 

Note that the chemical potential depends on temperature through the relation; $\int_0^{\infty} D(\epsilon) f_{\mbox{\scriptsize FD}}(\epsilon, T) \; d\epsilon = n_0$, where $n_0$ is the number of electrons in a system. Instead of giving $n_0$, the chemical potential at the ground state $\mu_0$, which corresponds to the Fermi energy in the ``doped'' system, is specified in this work. (Hereinafter, the Fermi energy in the ``non-doped'' and ``doped'' systems are represented as $\epsilon_F$ and $\mu_0$, respectively.)

\begin{figure}
\includegraphics[scale=0.115]{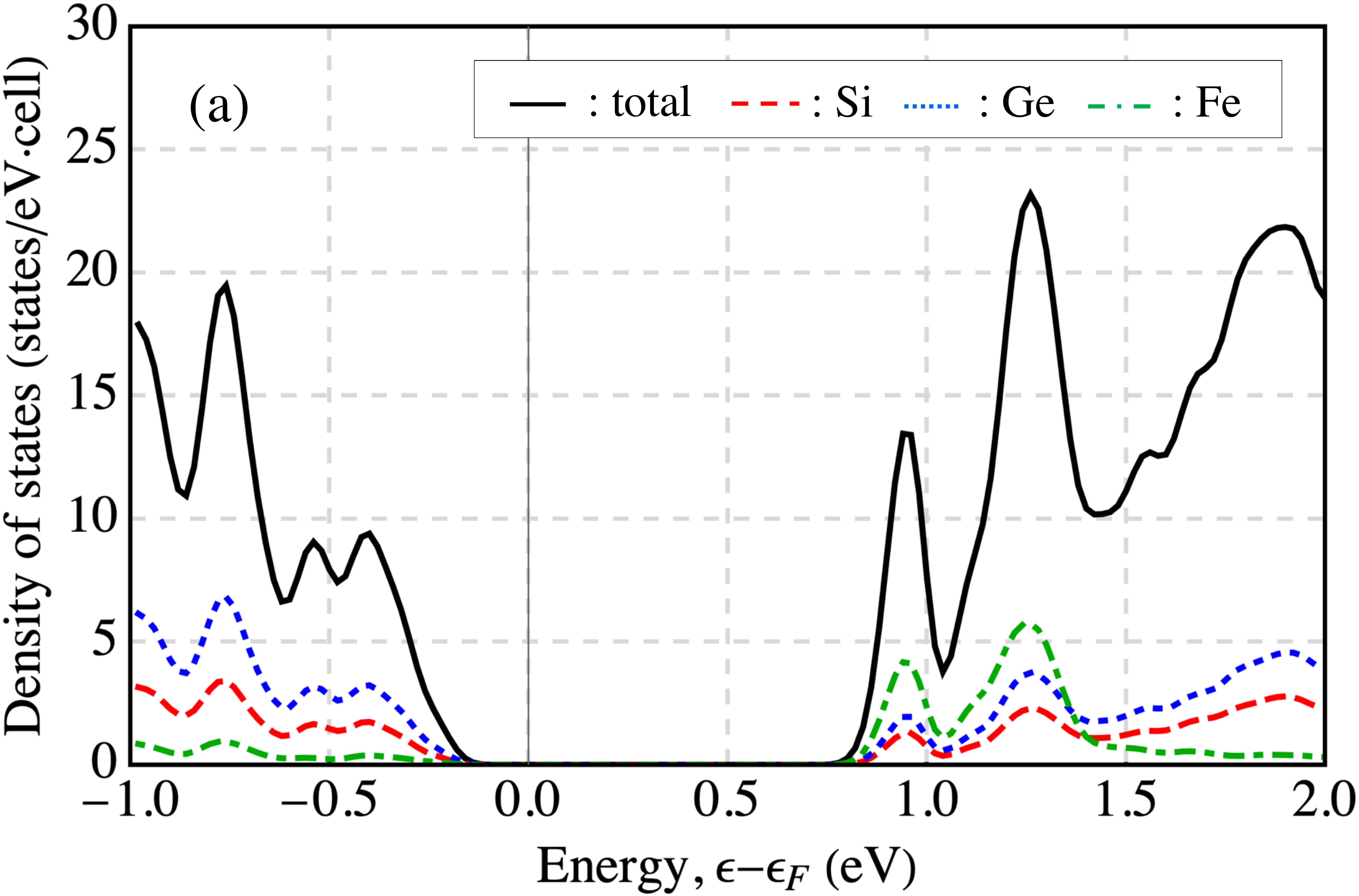}
\includegraphics[scale=0.115]{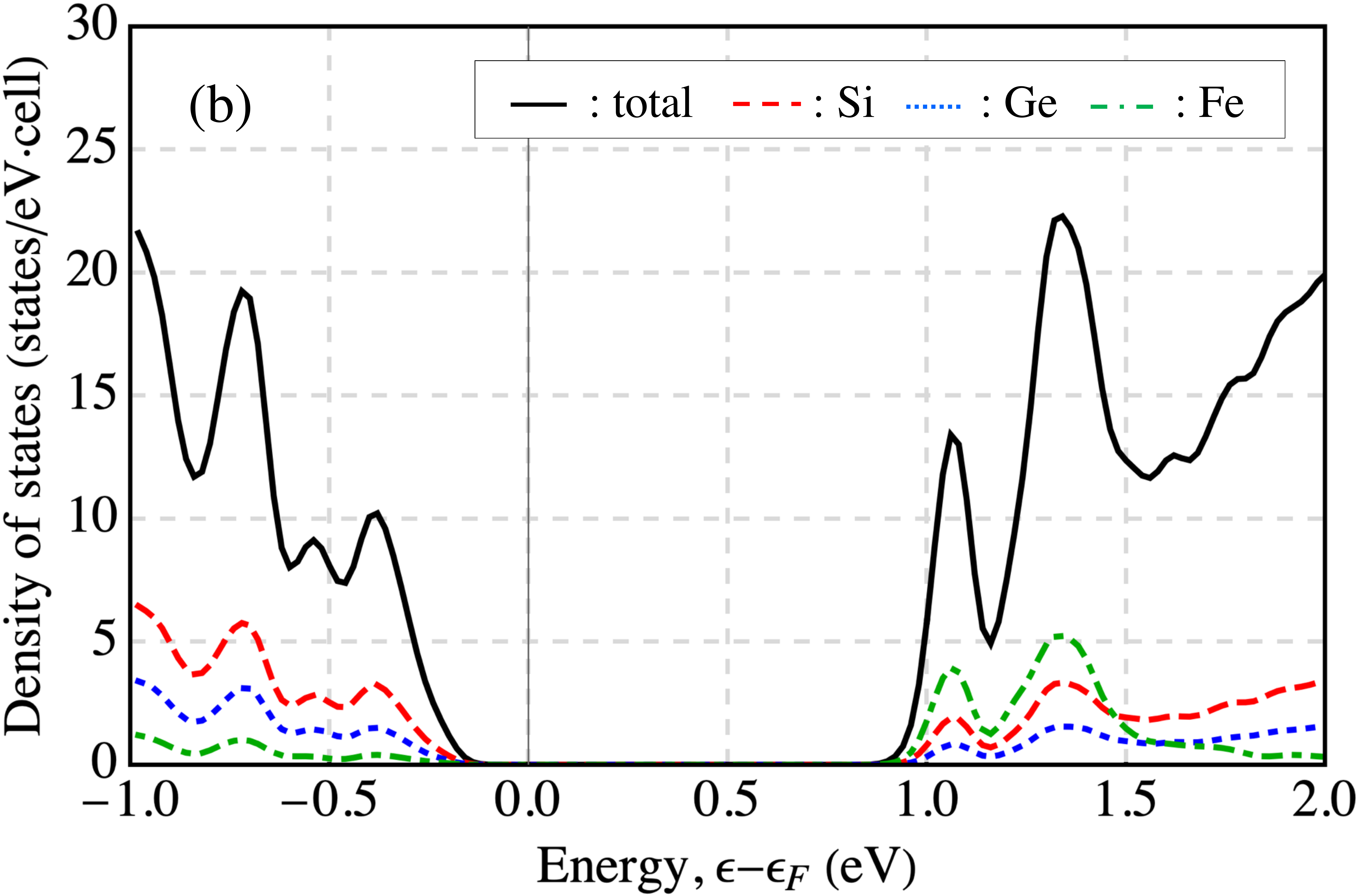}
\caption{\label{fig:dos_Fe_Si_Ge} Calculated electronic density of states in the (a) Si$_{0.500}$Ge$_{0.484}$Fe$_{0.016}$ and (b) Si$_{0.781}$Ge$_{0.203}$Fe$_{0.016}$ alloys. The black solid line is the total density of states, and the red dash, blue dotted, and green dash-dotted lines indicate the partial density of states of Si, Ge, and Fe, respectively. }
\end{figure}

The calculated electronic density of states in the Si$_{0.500}$Ge$_{0.484}$Fe$_{0.016}$ and Si$_{0.781}$Ge$_{0.203}$Fe$_{0.016}$ alloys are, respectively, shown in Figs.\;\ref{fig:dos_Fe_Si_Ge}\;(a) and (b), where their partial density of states of Si, Ge, and Fe are also presented. From these results, one can see that there are two strong peaks mainly originated from the Fe-doping at the edge of the conduction band for both the compositions. A similar impurity state has been reported in the Si$_{0.993}$Fe$_{0.007}$ (or Si$_{143}$Fe$_1$) alloy system \cite{delime2019large}, where its electronic density of states was calculated using density functional theory with the GGA/PBE functional. 

\begin{figure}
\includegraphics[scale=0.16]{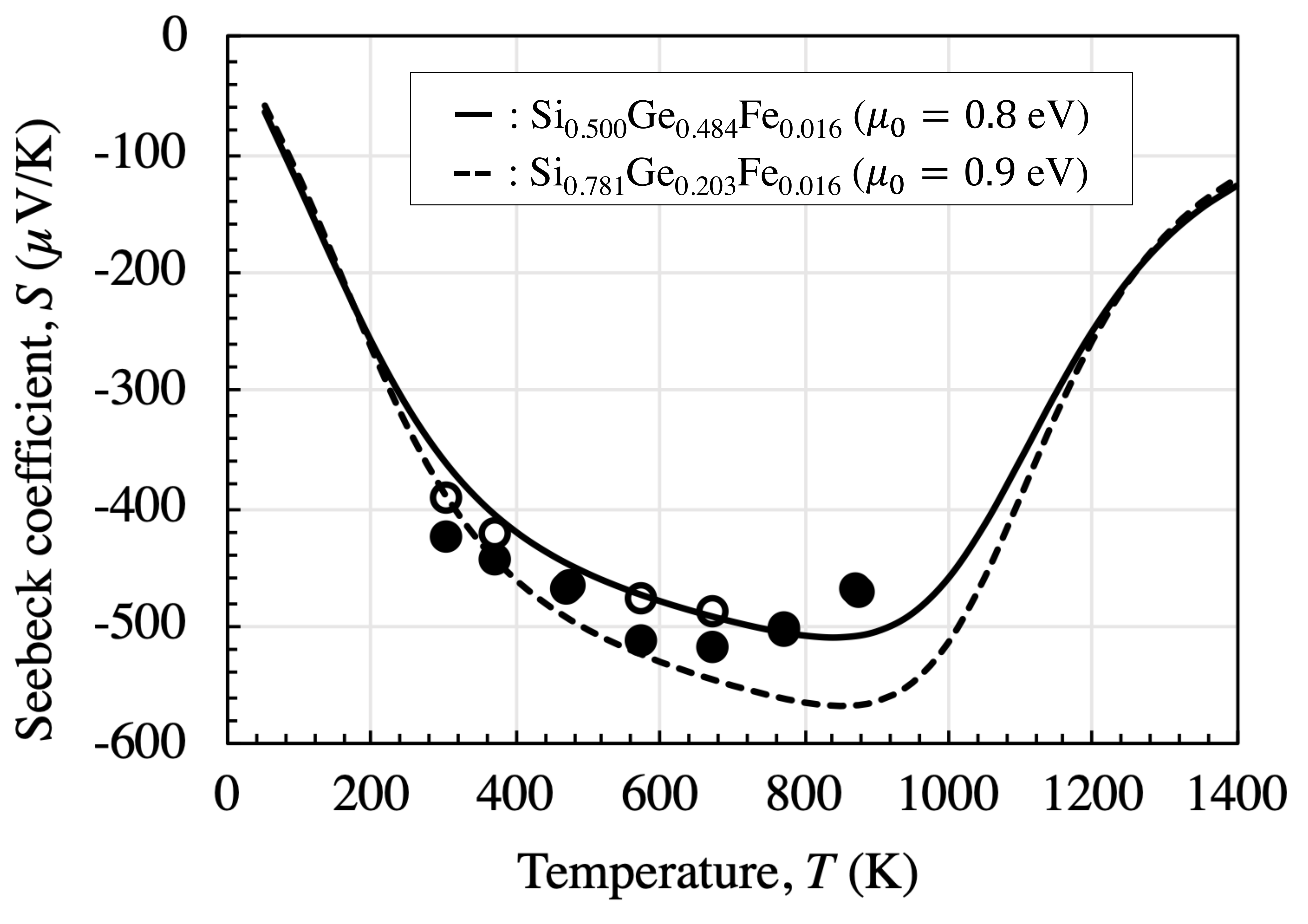}
\caption{\label{fig:seebeck_Fe_Si_Ge} Calculated Seebeck coefficients in the Si$_{0.500}$Ge$_{0.484}$Fe$_{0.016}$ (solid line) and Si$_{0.781}$Ge$_{0.203}$Fe$_{0.016}$ (broken line) alloy systems. The experimental data measured in the nanostructured Si$_{0.55}$Ge$_{0.35}$P$_{0.10}$Fe$_{0.01}$ sample for the heating/cooling condition are also provided as open/filled circles \cite{delime2019large}. }
\end{figure}

The Seebeck coefficients calculated from Eq.\;\eqref{eq:seebeck_approximate} using the total density of states shown in Fig.\;\ref{fig:dos_Fe_Si_Ge} are presented in Fig.\;\ref{fig:seebeck_Fe_Si_Ge}. Here, the Fermi energies (or chemical potentials at the ground state), $\mu_0$, are set to the bottom of the conduction band. From Fig.\;\ref{fig:seebeck_Fe_Si_Ge}, the Seebeck coefficient of the Si$_{0.781}$Ge$_{0.203}$Fe$_{0.016}$ alloy shows a larger value compared to that of the Si$_{0.500}$Ge$_{0.484}$Fe$_{0.016}$ alloy. This is because the former composition has a larger band gap compared to the latter one, as can be seen in Fig.\;\ref{fig:dos_Fe_Si_Ge}. The experimental data for the nanostructured Si$_{0.55}$Ge$_{0.35}$P$_{0.10}$Fe$_{0.01}$ sample are also shown in Fig.\;\ref{fig:seebeck_Fe_Si_Ge}. One can see that the calculated Seebeck coefficients are quite close to the experimental data \cite{delime2019large}. 

\begin{figure}
\includegraphics[scale=0.133]{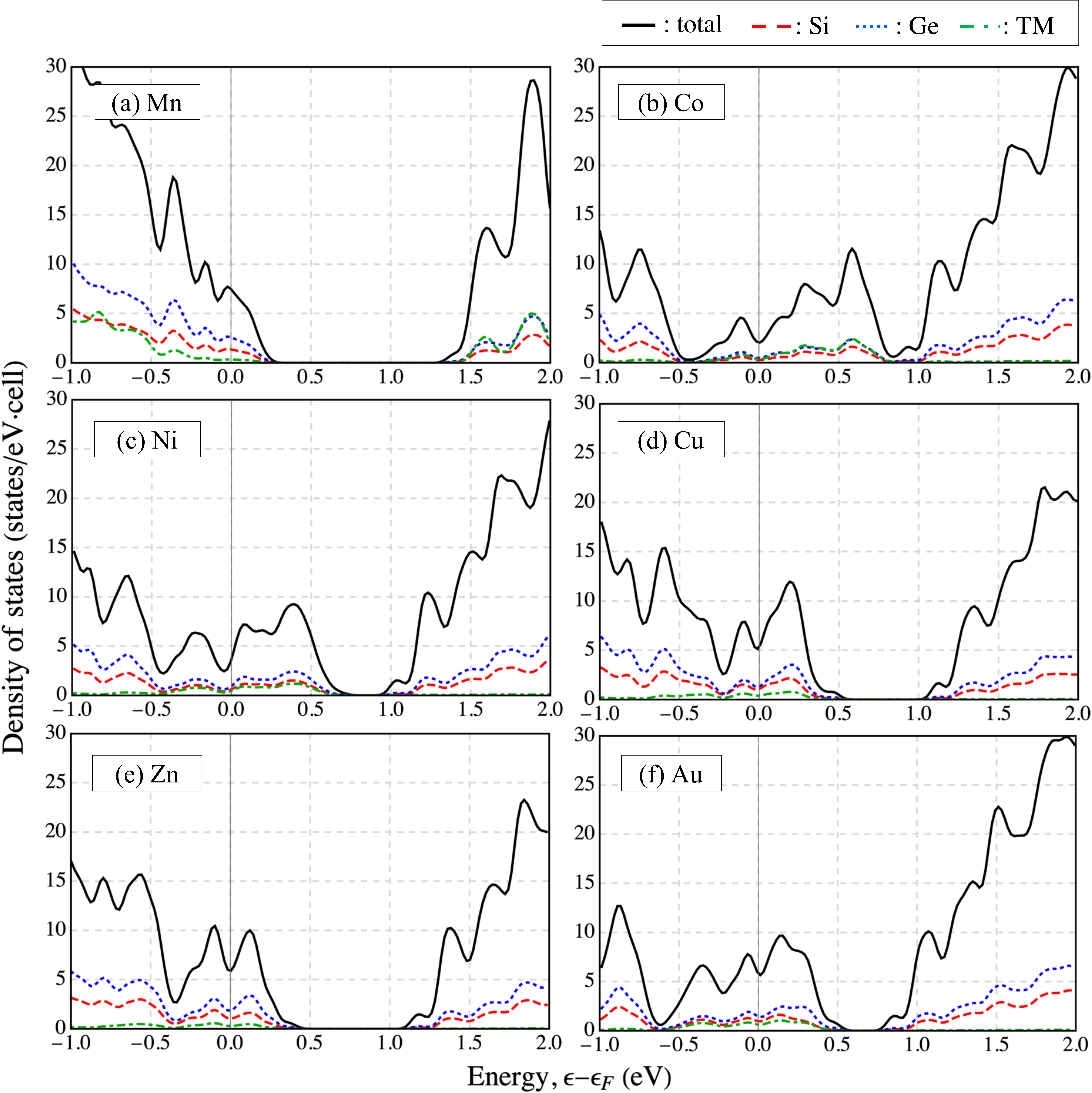}
\caption{\label{fig:dos_Si32Ge31TM1} Calculated electronic density of states in the Si$_{0.500}$Ge$_{0.484}$TM$_{0.016}$ alloy systems, where (a) Mn-, (b) Co-, (c) Ni-, (d) Cu-, (e) Zn-, and (f) Au-doped systems. The black solid lines are the total density of states, and the red dash, blue dotted, and green dash-dotted lines indicate the partial density of states of Si, Ge, and TM, respectively. }
\end{figure}

Next, electronic density of states and Seebeck coefficient in other transition-metal-doped systems are presented. The calculated electronic density of states in the Si$_{0.500}$Ge$_{0.484}$TM$_{0.016}$ alloy systems (TM=Mn, Co, Ni, Cu, Zn, and Au) are shown in Fig.\;\ref{fig:dos_Si32Ge31TM1}. From Fig.\;\ref{fig:dos_Si32Ge31TM1}, the impurity state originating from the TM-doping can be seen at the bottom of the conduction band in the Mn-doped system, whereas those in other TM-doped systems are at the top of the valence band. 

To obtain a large Seebeck coefficient, it is important to have not only a sharp peak at the end of either valence or conduction band, but also a large band gap (as can be seen in the Fe-doped system, Fig.\;\ref{fig:dos_Fe_Si_Ge}). Since the band gap in the Mn-doped system is larger than that in the Fe-doped one (see Fig.\;\ref{fig:dos_Fe_Si_Ge}\;(a)) and there are large impurity states, it is expected that the Mn-doped system has a larger Seebeck coefficient than that of the Fe-doped system. 

\begin{figure}
\includegraphics[scale=0.16]{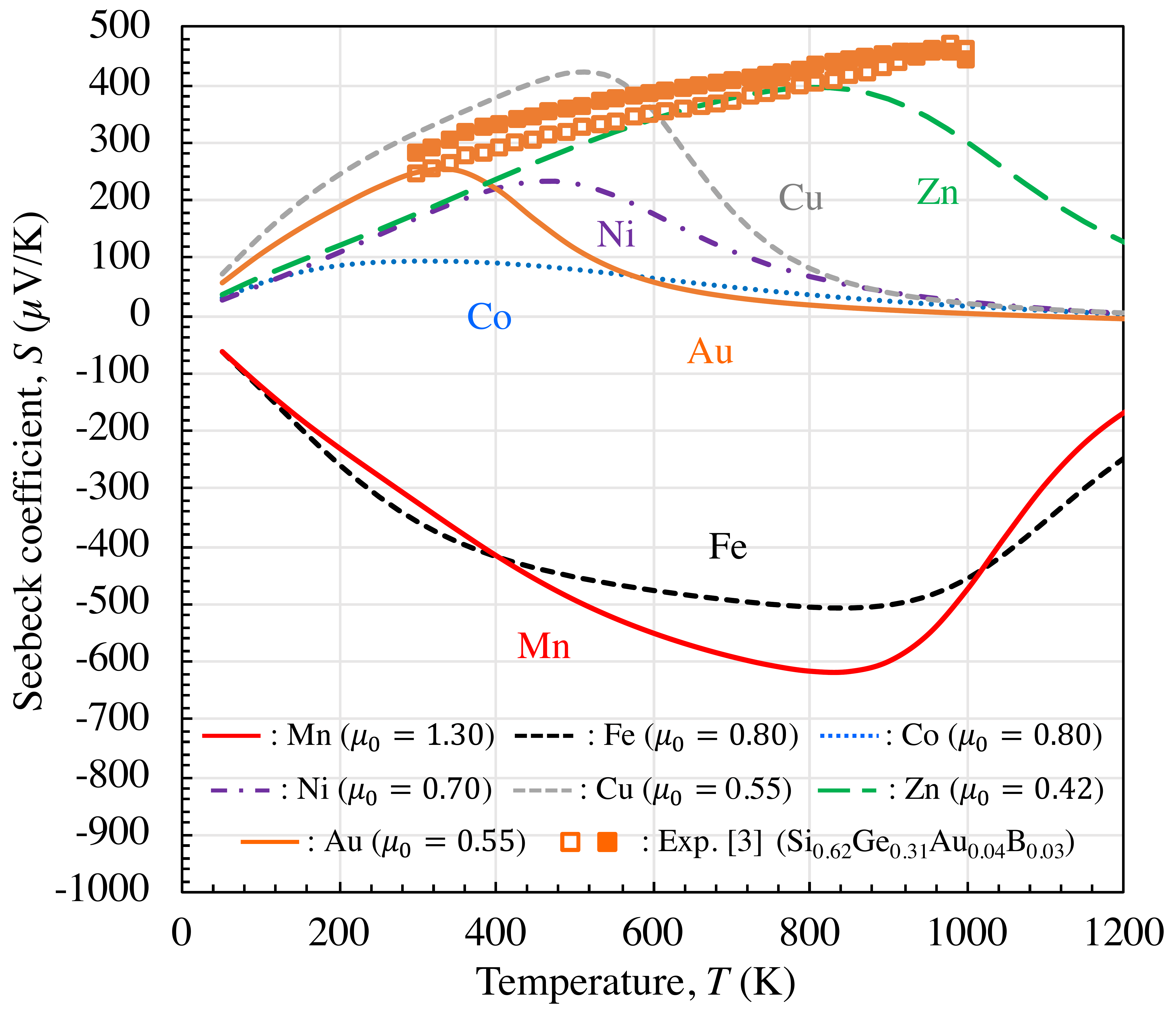}
\caption{\label{fig:seebeck_TM_Si_Ge} Calculated Seebeck coefficients in the Si$_{0.500}$Ge$_{0.484}$TM$_{0.016}$ alloys (TM=Mn, Fe, Co, Ni, Cu, Zn, and Au). The Fermi energies used in the calculations, $\mu_0$, are provided in the legend (the unit is eV). The experimental data measured in the nanostructured Si$_{0.62}$Ge$_{0.31}$Au$_{0.04}$B$_{0.03}$ sample for the heating/cooling condition are shown together as open/filled orange squares \cite{omprakash2019au}.}
\end{figure}

The calculated Seebeck coefficient in the TM-doped systems are shown in Fig.\;\ref{fig:seebeck_TM_Si_Ge}, where the result of the Fe-doped system (the same result shown in Fig.\;\ref{fig:seebeck_Fe_Si_Ge}) is presented as well. Depending on the value of the Fermi energy, $\mu_0$, the TM-doped systems show either $p$- or $n$-type thermoelectric characteristics. Here, the $\mu_0$ is set to either the end of valence or conduction band depending on the position of the impurity states. From Fig.\;\ref{fig:seebeck_TM_Si_Ge}, the magnitude of the Seebeck coefficients of the Co-, Ni-, Cu-, Zn-, and Au-doped systems are smaller than that of Fe-doped system, but the Mn-doped system shows larger Seebeck coefficient than that of the Fe-doped system at high temperatures ($400\sim 1000$\;K), as expected from their calculated electronic density of states. 

The experimentally measured Seebeck coefficients in the nanostructured Si$_{0.62}$Ge$_{0.31}$Au$_{0.04}$B$_{0.03}$ sample \cite{omprakash2019au} are also shown in Fig.\;\ref{fig:seebeck_TM_Si_Ge}. Compared to the experimental data, the calculated Seebeck coefficients in the Au-doped system are significantly underestimated at high temperatures. The main reason of this discrepancy would be the existence of a secondary phase in the measured sample \cite{omprakash2019au}. In addition, the difference in the alloy compositions between the calculation and experiments is also considered to be the cause of the discrepancy. Since band gap of Si is larger than that of Ge, it is expected that band gap is increased by increasing the fraction of Si in the calculated Au-doped system, which will result in the increase of Seebeck coefficient at high temperatures. 

From the calculated Seebeck coefficients, it is expected that the Mn-doping is better than the Fe-doping for the Si--Ge alloy system. However, the $ZT$ value depends not only on the Seebeck coefficient but also on the electrical resistivity, $\rho$, as $ZT=S^2 T/\rho \kappa$, where $\kappa$ is the thermal conductivity. In general, the electrical resistivity is increased with the band gap. Thus, the electrical resistivity (as well as the thermal conductivity) needs to be evaluated to conclude whether the Mn-doped system is better thermoelectric materials than the Fe-doped system or not. 

Note that it is assumed that the doped transition metals occupy substitutional sites, but it is uncertain whether they are located at substitutional or interstitial sites. To make clear their preferred sites from density functional theory, their formation energies need to be calculated and compared. However, this is beyond the scope of this paper and is left for future work.

In conclusion, the electronic density of states and Seebeck coefficient in the TM-doped Si--Ge systems (TM=Mn, Fe, Co, Ni, Cu, Zn, and Au) were investigated from the first-principles calculations with a hybrid functional (HSE06) using disordered configurations prepared based on the SQS. The impurity states in the Fe-doped Si--Ge systems were successfully produced, and the reported large Seebeck coefficients in the nanostructured Si$_{0.55}$Ge$_{0.35}$P$_{0.10}$Fe$_{0.01}$ sample were quantitatively reproduced from the computed electronic density of states. Using the same methodology, the electronic density of state and Seebeck coefficient of other TM-doped Si--Ge systems (TM=Mn, Co, Ni, Cu, Zn, and Au) were calculated. It was found that Mn-doping produces strong impurity states at the bottom of conduction band, and the See beck coefficient is larger than that of the Fe-doped system at high temperatures. Thus, Mn is considered to be a better dopant for the Si--Ge systems from the perspective of the Seebeck coefficient.

\section*{Acknowledgement}
This work was supported by Japan Science and Technology Agency (JST) CREST, Grant Number: JPMJCR1812.

\bibliographystyle{apsrev4-2}

\bibliography{ref}

\end{document}